\documentclass[aip,apl,reprint,superscriptaddress]{revtex4-1}
\usepackage[T1]{fontenc}
\usepackage{siunitx}
\usepackage{times}
\usepackage{float}
\usepackage{graphicx,color}
\usepackage{amsfonts,amsmath,amssymb,amsbsy}
\usepackage[colorlinks=true, linkcolor=blue, citecolor=red, urlcolor=magenta]{hyperref}

\def\blue#1{\textcolor{blue}{#1}}
\def\emph#1{\textcolor{red}{#1}}
\begin{document}

\title{Effect of Notch Structure on  Magnetic Domain Movement in Planar Nanowires}

\author{Hua Ling}
\thanks{These authors contributed equally to this work.}
\affiliation{Spintronics and Nanodevice laboratory, Department of Electronic Engineering,
University of York, York YO10 5DD, United Kingdom}
\affiliation{York-Nanjing International Center of Spintronics (YNICS), Nanjing University, Nanjing 210093, China}

\author{Junlin Wang}
\thanks{These authors contributed equally to this work.}
\affiliation{Spintronics and Nanodevice laboratory, Department of Electronic Engineering,
University of York, York YO10 5DD, United Kingdom}
\affiliation{York-Nanjing International Center of Spintronics (YNICS), Nanjing University, Nanjing 210093, China}

\author{Xianyang Lu}
\affiliation{Department of Physics, University of York, York YO10 5DD, United Kingdom}
\affiliation{York-Nanjing International Center of Spintronics (YNICS), Nanjing University, Nanjing 210093, China}

\author{Junran Zhang}
\affiliation{York-Nanjing International Center of Spintronics (YNICS), Nanjing University, Nanjing 210093, China}

\author{Li Chen}
\affiliation{Faculty of Engineering, University of Leeds, Woodhouse Lane, Leeds LS2 9JT, United Kingdom}

\author{Christopher Reardon}
\affiliation{Department of Physics, University of York, York YO10 5DD, United Kingdom}

\author{Jason Zhang}
\affiliation{Department of Physics, University of York, York YO10 5DD, United Kingdom}

\author{Yichuan Wang}
\affiliation{Department of Physics, University of York, York YO10 5DD, United Kingdom}
\affiliation{York-Nanjing International Center of Spintronics (YNICS), Nanjing University, Nanjing 210093, China}

\author{Yu Yan}
\affiliation{Spintronics and Nanodevice laboratory, Department of Electronic Engineering, University of York, York YO10 5DD, United Kingdom}
\affiliation{York-Nanjing International Center of Spintronics (YNICS), Nanjing University, Nanjing 210093, China}

\author{Jing Wu}
\email{jing.wu@york.ac.uk}
\affiliation{Department of Physics, University of York, York YO10 5DD, United Kingdom}
\affiliation{York-Nanjing International Center of Spintronics (YNICS), Nanjing University, Nanjing 210093, China}

\author{Yongbing Xu}
\email{yongbing.xu@york.ac.uk}
\affiliation{Spintronics and Nanodevice laboratory, Department of Electronic Engineering,
University of York, York YO10 5DD, United Kingdom}
\affiliation{York-Nanjing International Center of Spintronics (YNICS), Nanjing University, Nanjing 210093, China}

\begin{abstract}
We present the direct observation of magnetic domain motion in permalloy nanowires with notches using a wide-field Kerr microscopy technique. The domain wall motion can be modulated by the size and shape of the notch structure in the nanowires. It is demonstrated that the coercive fields can be tuned by modulating the notches.  The experimental results are consistent with the micro-magnetic simulation results. The relationship between the notch angle and the domain nuclease are also studied. This work is useful for the design and development of the notch-based spintronic devices.
 
\end{abstract}

\date{\today}
\maketitle


Recently, the study of domain walls (DWs) motion along planar nanowires has attracted many scientific interests because of the potential to be used in the logic and racetrack memory devices.\cite{allwood2005magnetic,parkin2008magnetic} A transverse or vortex domain wall propagate in a magnetic nanowire depend on the ratio of the width and thickness.\cite{nakatani2005head} A unique assembled nanowire structure such as a notch or a protrusion create pinning sites for DWs.\cite{himeno2005propagation,petit2008domain,lepadatu2007current} It is the core of the development of spintronic devices. Magneto-optic Kerr effect (MOKE) is one of the most powerful techniques to study domain wall motion. The MOKE  measerements without imaging\cite{cormier2008high} can provide observation of the DWs motion which has been used to examine in different planar permalloy nanostructure, such as, L-shape nanowire,\cite{allwood2002shifted} rings,\cite{moore2005magnetization} elliptical\cite{sebastian2011magneto} and pad-notch nanowire.\cite{wang2017magnetic,faulkner2003controlled,faulkner2004artificial,burn2014control,
bogart2009dependence,zhu2012depinning,xiong2001magnetic,petracic2007systematic,
bogart2008effect,faulkner2008tuning} The shifted hysteresis loops can be obtained from L-shape nanowires. The study of a combination of MOKE and simulations gave an ideal explanation of rings magnetic reversal of magnetization switching\cite{moore2005magnetization}. Additionally, the study of different sizes and orientation in permalloy elliptical shape demonstrated the relationship between the shape anisotropy and the coercivity of devices\cite{zhu2012depinning}. Most of the pad-notch nanowire structure studies display that the shape anisotropy make the domain nuclear at the larger structure and pinned at a junction between the large structure and wire\cite{petit2008domain}. And the location of the domain can be pinned at the notch structure of the device. However, those studies are based on analysing the regulations of hysteresis loops and comparing with micro-magnetic simulation. In this letter, we report the direct observation of the DWs propagation process  by the MOKE imageing technique. It provides magnetization reversal images which demonstate the DWs motion in the nanowires. The 3D notch structure can be used for the magnetic ratchet memory device\cite{franken2012shift,reichhardt2017ratchet,wang2017controllable} and which shows the potential use as next generation spintronics device designing.




The micro-magnetic simulations are performed using the standard micro-magnetic simulator Object Oriented MicroMagnetic Framework software,\cite{donahue2010oommf} which includes domain wall motion by applying the Landau-Lifshitz-Gilbert equation.\cite{liu2007spin} 

\begin{equation}
\frac{d\boldsymbol{M}}{dt}=-\left|\gamma\right|\boldsymbol{M}\times\boldsymbol{H}_{\text{eff}}+\frac{\alpha}{M_{\text{S}}}\left(\boldsymbol{M}\times\frac{d\boldsymbol{M}}{dt}\right),
\label{LLG}
\end{equation}

where $\boldsymbol{M}$ is the magnetization of the magnetic layer, $M_{\text{S}}$ is the saturation magnetization, $\gamma$ is the Gilbert gyromagnetic ratio and $\alpha$ is the damping constant. $\boldsymbol{H}_{\text{eff}}$ is the effective field, which is derived from the magnetic energy density,

\begin{equation}
\boldsymbol{H}_{\text{eff}}=-\mu_{0}^{-1}\frac{\delta\varepsilon}{\delta\boldsymbol{M}},
\label{energy-density}
\end{equation}

where $\varepsilon$ contains the Heisenberg exchange, anisotropy, applied magnetic field, and demagnetization energy terms. The simulation parameters are set as saturation magnetization $M_{\rm{s}}=\SI{8.6e5}{A/m}$, exchange coefficient $A_{\rm{ex}}=\SI{1.3e-11}{J/m}$, Gilbert damping constant $\alpha = 0.01$. The cell size is chosen as $\SI{20}{nm}$ for the reasonable time.

\begin{figure}[t]
\centerline{\includegraphics[width=3.375in]{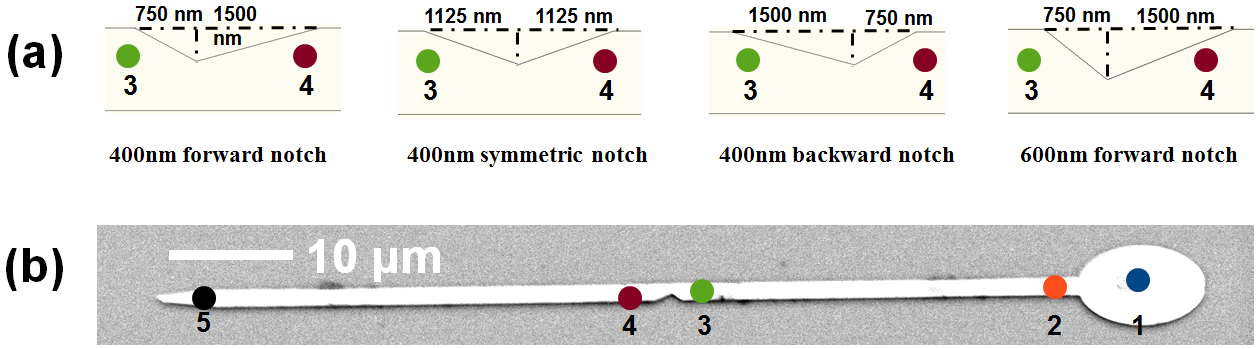}}
\caption{%
(a) Schematic configurations for four different triangles notch. The first three are $\SI{400}{nm}$ notch depth, and the last one is $\SI{600}{nm}$ notch depth. (b) SEM image for $\SI{400}{nm}$ forward notch shows the location of the detect position. The different numbers and colours represent different detecting position on nanowires.
}
\label{FIG1}
\end{figure}

In this report, we used Kerr image measurement and micro-magnetic simulation to study DWs pinning/depinning in a permalloy (Ni$_{20}$Fe$_{80}$) nanowire with different geometry of notch. The height of the triangles is $\SI{400}{nm}$ with forward, backward and symmetric configuration. The device with $\SI{600}{nm}$ depth forwarding notch structure is shown in Fig.~\ref{FIG1}(a). The width of the wires is $\SI{1}{\micro\meter}$ with a $\SI{6}{\micro\meter}\times\SI{8}{\micro\meter}$ nucleation pad at the end of nanowires. The nanowires are designed to be long enough with a total length as  $\SI{86}{\micro\meter}$ such that the regions of the nanowire before and after the pinning structure could be interrogated independently. The samples for measurement are fabricated by a combination of lift-off and e-beam lithography techniques and covered by $\SI{3}{nm}$ gold capping layers for anti-oxidation protection.

The magnetization behavior of individual nanowires is measured using a Kerr image microscopy which is built up based on a Kohler illumination\cite{schafer2007investigation} light path using an LED light source with a wavelength of $\SI{455}{nm}$. A high-resolution CCD camera captured polarisation dependent images showing the distribution of magnetization across a sample. By subtracting a domain-free background reference image from the magnetically saturated state, a clear micrograph of the domain pattern is obtained independently of topographic contrast. The quality of the observation is improved by averaging. However,  the permalloy  nano-structure is thin and small which makes the kerr signal  with a very weak value as  $0.4\%$. 


In simulation, parameters of the micro-magnetic models are the same as the samples using in the experiment, except if there is any defect caused by fabrications. All modelling will start from a randomly generated initial magnetic state and allowed to relax at 0 Oe until it stabilizes. Then the magnetic field will increase from $\SI{0}{Oe}$ to $\SI{1000}{Oe}$ in $\SI{10}{Oe}$ increment. After that, magnetic fields are reduced from the $\SI{1000}{Oe}$ to $\SI{-1000}{Oe}$ in $\SI{10}{Oe}$ decrements and back to $\SI{1000}{Oe}$ with $\SI{10}{Oe}$ increment.
\begin{figure}[t]
\centerline{\includegraphics[width=3.375in]{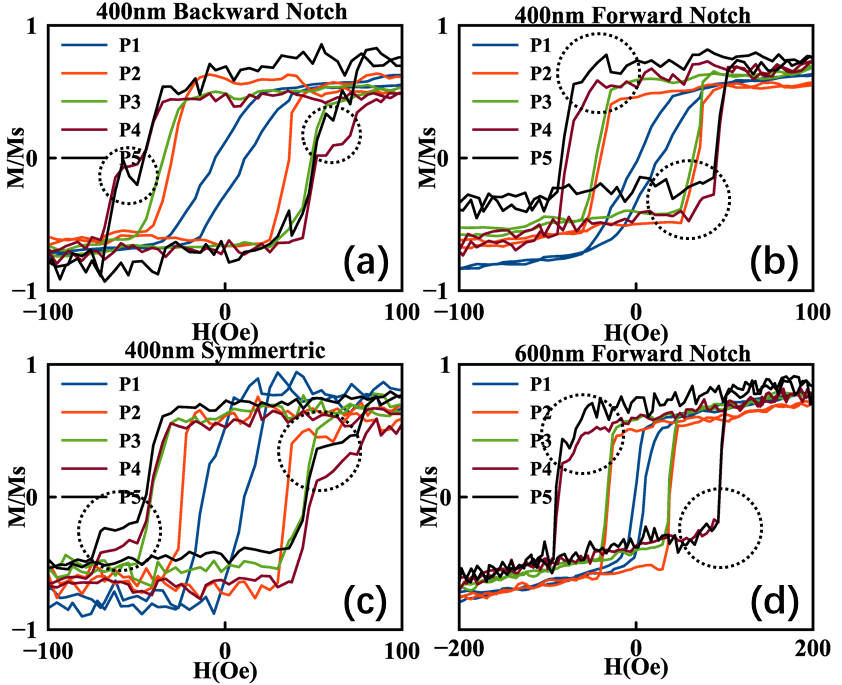}}
\caption{%
The hysteresis loops are achieved by wide-field Kerr microscope. The circle indicates double domain appearing external field. (a) Hysteresis loops of $\SI{400}{nm}$ backward notch configuration. (b) is hysteresis loops of $\SI{400}{nm}$ forward notch configuration. (c) Hysteresis loops of $\SI{400}{nm}$ symmetric notch configuration. (d) Hysteresis loops of 600 nm forward notch configuration.The positions are the location of detecting points shown in Fig.~\ref{FIG1}(b)and the Postion 1 - Position 5 are written as P1-P5. The configuration of the notch is displayed on top of each graph.
}
\label{FIG2}
\end{figure}

\begin{figure*}[!]
\centerline{\includegraphics[width=6.75in]{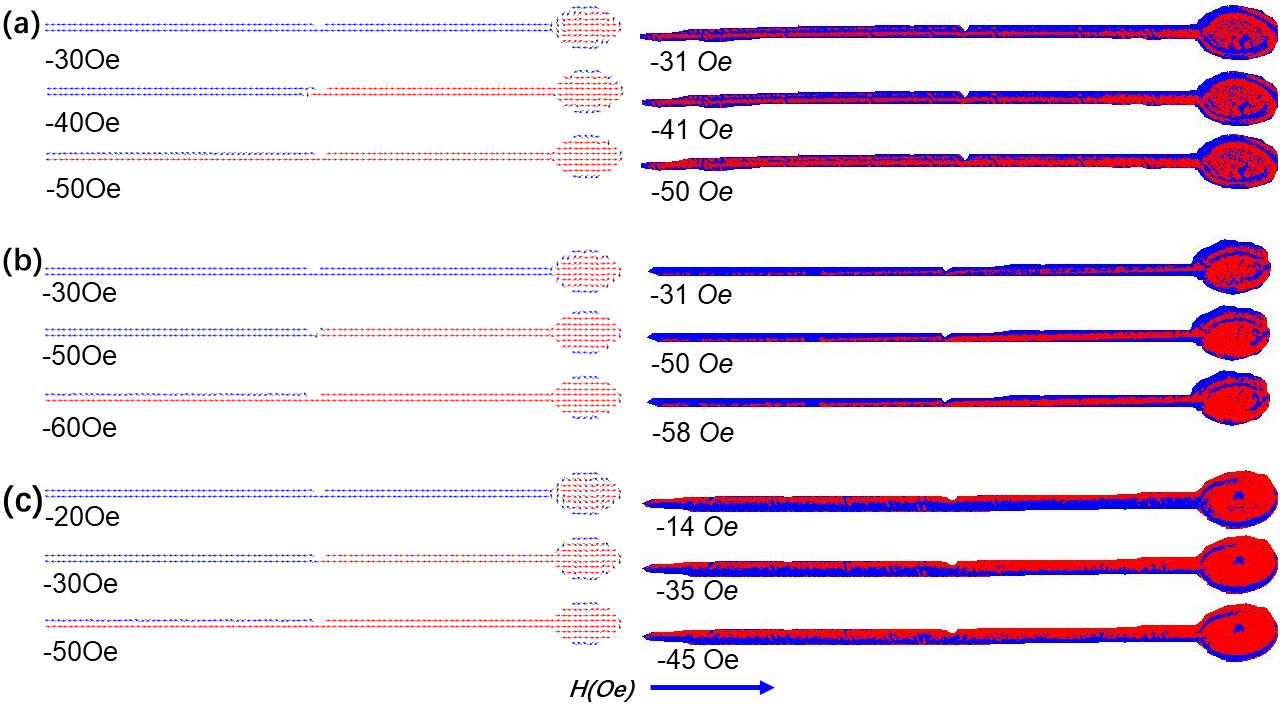}}
\caption{%
Comparison OOMMF results and between Kerr contrast images for $\SI{400}{nm}$ symmetric, backward and forward notch wire.  Simulation images are shown in the right panel and the equivalent phenomenon of contrast images are shown in the left panel and. The blue arrow and color show the positive magnetic field direction. (a)  the symmetric notch configuration at pad nucleation state, pinning state and depinning state respectively. (b)  the backward notch configuration at pad nucleation state, pinning state and depinning state respectively. (c)  the forward notch configuration at pad nucleation state, pinning state and depinning state respectively. 
}
\label{FIG3}
\end{figure*}


The hysteresis loops are shown in Fig.~\ref{FIG2} represent that increasing of distance from the nucleation pad cause the increasing of coercivity. Fig.~\ref{FIG2} (a)-(c) show that the coercivity of position 4(P4) and 5(P5) in the nanowires with different geometry of 400nm notch are similar. The P4 and P5 in the nanowires with a 400nm notch has a larger coercivity than the other position in the nanowire which are marked in Fig.~\ref{FIG2}. And from Fig.~\ref{FIG2} (d), it can be found that the coercivity of P4 ad P5 in the nanowire with $\SI{600}{nm}$ forward notch is much larger than the nanowires with $\SI{400}{nm}$ depth notch. The shape anisotropy of the nanowire gives an effect on the coercivity of the device. When the depth of the notch increasing, the pinning/depinning magnetic field increasing which means the domain wall is more difficult to pass the notch structure. In addition, the coercivity of P1-P3 in the $\SI{600}{nm}$ notch nanowire is similar to the nanowires with $\SI{400}{nm}$ notch. The coercivity of $\SI{600}{nm}$ notch nanowire in P4 and P5 is much larger than the coercivity in other positions in this case. Fig.~\ref{FIG2} (a) and (d) demonstrate that the depth of the notch gives deep effect on the pinning/depinning magnetic field of the nanowire. The Hysteresis loops of $\SI{600}{nm}$ depth symmetric notch configuration and $\SI{600}{nm}$ depth backward notch are also given in the supplementary information. \blue{(See Supplementary Information Fig. S1)}


The contrast images in Fig.~\ref{FIG3} show the typical domain pinning/depinning states behaviour of different structures. The domain dynamics in the nanowire with a $\SI{400}{nm}$ forward notch is shown in Fig.~\ref{FIG3} (a)-(c). When the applied magnetic field decreasing from 0 to $\SI{-31}{Oe}$ the domain will nucleates in the pad structure of the nanowire with symmetric notch. Fig.~\ref{FIG3} (a) shows that domain has been pinned at the notch structure with a applied magnetic field as $\SI{-41}{Oe}$. When the applied magnetic field increase to $\SI{-45}{Oe}$, the domain wall will pass the notch structure. Moreover from the hysteresis loops of the $\SI{400}{nm}$ symmetric nanowire in Fig.~\ref{FIG2} (b), it shows that the magnetization of the whole nanowire drop into reverse value. In the nanowire with $\SI{400}{nm}$ backward notch, the domain nucleates in the pad of the nanowire with an applied magnetic field is $\SI{-31}{Oe}$. The domain pass the notch structure when the applied magnetic field larger than $\SI{-60}{Oe}$. In the nanowire with $\SI{400}{nm}$ forward notch, the domain nucleates in the pad when the applied magnetic field is $\SI{-14}{Oe}$. And the depinning magnetic filed of the domain in the nanowire is $\SI{-45}{Oe}$. Comparing with nanowires with $\SI{400}{nm}$ symmetric notch and $\SI{400}{nm}$ backward notch, the nanowire with a forward notch has a smaller depinning magnetic field and the domain is easier nuclease in the pad of the nanowire. The performance of the nanowire depends on the gap between the domain nucleation magnetic field and domain depinning magnetic field. According to Fig.~\ref{FIG3} (a) - (c), it can be found that the nanowire with a $\SI{400}{nm}$ forward notch displays a best performance in the nanowires with $\SI{400}{nm}$ depth notch. On the other hand, the nanowire with a $\SI{400}{nm}$ backward has the largest depinning mangetic field which means the nanowire with backward notch is the most stable one. The simulation results are shown in the left of Fig.~\ref{FIG3}. The simulation results are good agreement with the experiment results except for the results in Fig.~\ref{FIG3} (c). In the simulation result, the domain pass the notch structure and then change to a magnetic vortex. The similar phenomenon also happen in the experiment. But the magnetic vortexes display anti-clockwise vortex in the micro-magnetic simulation and display clockwise vortex in the experiment. The domain configuration of $\SI{600}{nm}$ depth notch nanowires are given in the supplementary information. \blue{(See Supplementary Information Fig. S2-S4)}

\begin{figure}[t]
\centerline{\includegraphics[width=3.375in]{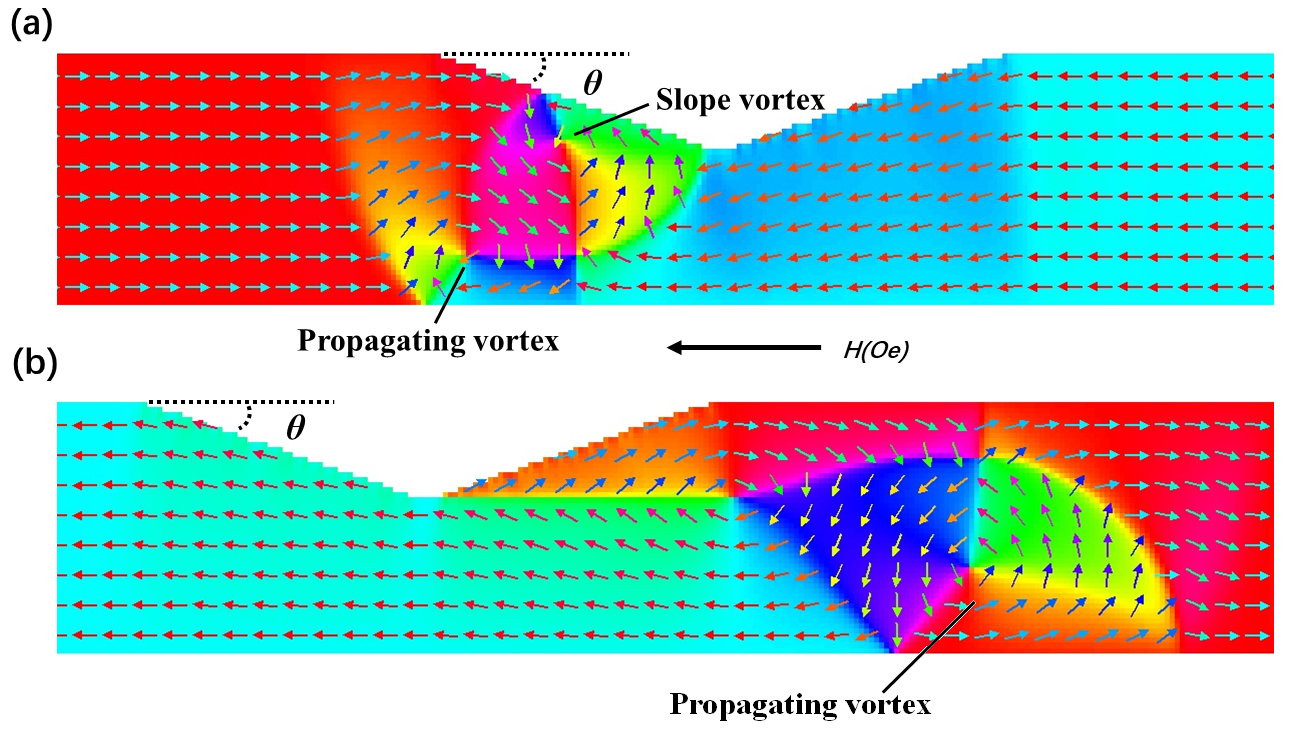}}
\caption{%
The magnetic simulation results on symmetric notched nanowire at notch area with a toward angle $\theta$. (a) The propagating vortex and slope vortex in the nanowire with an external field at before $\SI{-40}{Oe}$. (b) The propagating vortex with the external field at $\SI{-40}{Oe}$.
}
\label{FIG4}
\end{figure}

The jump in the circle is shown in Fig.~\ref{FIG2} indicates that there a double-domain structure display near the notch. The phenomenon can be seen as a horizontal domain wall separating a straight wire to top and bottom opposite domain. The phenomenon can be observed by Kerr image microscopy contrast image which is shown in Fig.~\ref{FIG3} with the domain depinning at the notch structure. The micro-magnetic simulation results in Fig. ~\ref{FIG4} demonstrate the details of the double domain structure. When the domains moving in the nanowire, two large vortex domain walls involved as indicated in Fig. \ref{FIG4}(b). One of the vortex DW is nucleated by notch structure and another one comes from along the nanowire. The interaction between this two vortex DWs caused a delay of pinning field. From Fig. \ref{FIG4}(a)-(b), it can found that the propagating vortex pass the notch structure when the external field is large enough. And the slope vortex disappear near the notch structure. After the vortex DWs passed the notch, the propagating vortex domain wall keeps propagating to the end of the wire. On the other hand a new vortex nucleates in the right of the notch structure which depend on the angle $\theta$ between the edge and the notch structure. A smaller angle $\theta$ can nucleates a large slope vortex and the size of the propagating vortex depend on the wire width. The large slope vortex slowed down the speed of the vortex propagating.  This phenomenon is shown in Fig.~\ref{FIG4}(a). The propagating vortex domain wall continued moving toward to the end of the nanowire and the another slope vortex DWs nucleate by other side of notch, then the double domain structure is generated. At other side of notch, the pad keeps nucleating slope vortex DWs and the vortex DWs will propagate to end of the wire until the full wire is saturated at saturation field.


In conclusion, the experiment results are in good agreement with the micro-magnetic simulation results. And the simulation shows the details of the DWs motion. The relationship between the pinning/depinning field and the notch structure is determined by the micro-magnetic simulation results and experiment results. The contrast images provided the details of the DWs motion and the hysteresis loops observed the area magnetization moment. From the experiment results, the hysteresis loop can be manipulated by adjusting the triangle shape of the notch. The depth of the notch also can affect the coercivity of the nanowire. And the size of the vortex domain can be modulated by the toward angle $\theta$ of the notch structure. The large toward angle $\theta$ can make the propagating domain pass the notch structure easily. Our results are useful for the design and development of the notch-based spintronic devices. \\

See \blue{supplementary material} for hysteresis loops of $\SI{600}{nm}$ depth notch and domain configuration of 600 nm depth notch.

\section*{Acknowledgment}

This work was supported by State Key Program for Basic Research of China (Grant No.  2014CB921101, 2016YFA0300803 ), NSFC (Grants No. 61427812, 11574137), Jiangsu NSF (BK20140054), Jiangsu Shuangchuang Team Program and the UK EPSRC (EP/G010064/1).

\bibliography{References}

\end{document}